\documentclass[11pt]{article}
\usepackage{epsfig}
\usepackage{graphicx}
\newcommand{\be}{\begin{equation}}
\newcommand{\bE}{{\bf e}}
\newcommand{\ee}{\end{equation}}

\newcommand{\bx}{{\bf x}}

\newcommand{\bF}{{\bf f}}

\newcommand{\eq}[1]{(\ref{#1})}
\newcommand{\fig}[1]{Fig.~\ref{#1}}

\newcommand{\lab}[1]{\protect\label{#1}}

\begin{document}

\title{\bf  A continuum description of the\\ energetics  and evolution of
stepped surfaces\\ in strained nanostructures}

\author{V.~B.~Shenoy and L.~B.~Freund \\
Division of Engineering, Brown University, Providence, RI 02912}

{\date{  \small\today}}

\maketitle

\begin{abstract}

\narrower{\small As a departure from  existing continuum approaches for
describing the stability and evolution of surfaces of crystalline materials,
this article provides a description of surface evolution based on the physics of
the main feature imposed by the discrete nature of the material, namely,
crystallographic surface steps.  It is shown that the formation energy of
surface steps depends on the sign of extensional strain of the crystal surface,
and this behavior plays a crucial role in surface evolution.  The nature of this
dependence implies that there is no energetic barrier to nucleation of islands
on the growth surface during deposition, and that island faces tend toward
natural orientations which have no counterpart in unstrained materials.  This
behavior is expressed in terms of a small number of parameters that can be
estimated through atomistic analysis of stepped surfaces. The continuum
framework developed is then applied to study the time evolution of surface shape
of an epitaxial film being deposited onto a substrate.  The kinetic equation for
mass transport is enforced in a weak form by means of a variational formulation.
It is found that islands form without nucleation barriers and they evolve to
shapes with natural surface orientations. The implications of the calculations
are shown to be consistent with the behavior observed during deposition of
semiconductor materials in recently reported experiments. Finally, it is
verified that the predictions of the continuum model are essentially the same as
those of the discrete step model for an isolated strained island. The
development in this article is limited to two-dimensional plane strain
deformation to keep the arguments transparent, but this is not a fundamental
limitation of the approach.}
\medskip

\end{abstract}

\setlength{\baselineskip}{16pt}

\section{Introduction}

Self-organized semiconductor nanostructures hold the promise for manufacture of
micro-electronic devices with unprecedented performance characteristics. Because
of the close connection between size and electronic characteristics, these
structures can be tuned to very specific requirements. Potential applications
include field effect transistors, quantum memory devices, and solid-state
lasers.

Strain-driven nucleation, growth and coarsening of epitaxial islands  offers a
versatile approach to manufacturing nanoscale devices. To exploit this
phenomenon in manufacturing, a fundamental understanding of the role of strain,
surface energies, and the kinetics of transport on the formation and evolution
of material structures is required.

Surface energy is an important concept in considering the evolution of
microstructure in small-scale material systems. The free energy of the bounding
surface of a crystal is a macroscopic quantity representing, in some sense, the
net work that had to be done to create that surface. This energy is distributed
over a specific mathematical surface that approximates the physical boundary
between the crystal and its surroundings. From this  definition of surface
energy it is clear that the reference level for energy of a free surface is the
state of the material on that same crystallographic surface when it is embedded
deep within a perfect crystal.  A number of analyses of surface stability and
evolution have been reported.  Asaro and Tiller (1972), Grinfeld (1986) and
Srolovitz (1989) independently showed that the flat surface of a stressed solid
under two-dimensional plane strain conditions is unstable if the surface energy
is independent of surface orientation.  The same physical model was extended to
three dimensions, but restricted to small amplitude surface fluctuations, by
Freund (1995).  A numerical method for handling large amplitude surface
fluctuations was developed by Zhang and Bower (1998).  Gao and Nix (1999)
adopted the same physical framework for describing the breakup of an unstable
film into islands. In all of these studies, the surface energy was assumed to be
isotropic, independent of surface strain and to have no connection to the
discrete nature of the crystalline materials being modeled.  This study is
focused on a departure from this point of view.

\begin{figure}[h]
\begin{center}
\includegraphics[height=12.0cm,width=12.0cm]{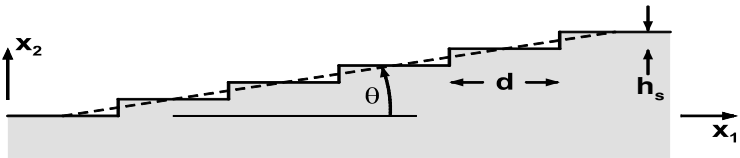}
\vspace{-9.0cm}
\parbox{1.0cm}{\ }\
\parbox{14cm}{
\caption{\lab{vicinalsurf}Geometry of a vicinal surface that makes an angle
of $\theta$ with a high symmetry direction. The spacing $d$ between the steps
is related to the step-height $h_s$ through the relation $d = h_s/\tan{\theta}$.
The dashed line shows the macroscopic surface orientation.}}
\end{center}
\label{fig1}
\end{figure}

The value of surface energy per unit area of a given crystallographic surface
orientation is determined by the fine scale structure of that surface.  For a
high symmetry orientation in a crystal, such as a \{100\} surface of a cubic
crystal, the surface is atomically flat. For other orientations close to this
surface, the structure usually consists of flat terraces with well defined local
surface energies, separated by atomic scale ledges or steps as illustrated
schematically in \fig{fig1}.  The  steps alter the macroscopic surface energy by
an amount corresponding to their energy of formation in the configuration
relevant to the structure.  Furthermore, the energies of individual features and
the interactions of these features are influenced by the presence of strain in
the crystal.   Macroscopically, the surface is assumed to be smooth but to have
a local surface energy at a point on the surface determined by the orientation
of the tangent plane at that point and the level of elastic strain in the
crystal. A quantitative interpretation of the free energy of a strained crystal
in terms of the physics of crystallographic steps is developed in the sections
that follow.  The driving force for alterations in surface morphology is the
surface chemical potential, defined in terms of variations of system free energy
with surface shape. The surface chemical potential of a stepped surface is also
derived, and implications for morphology evolution by surface diffusion are
examined. The actual values of the parameters involved in the characterization
can be determined only through atomistic simulation or experiment; such work
focused on the SiGe material systems is being carried out in parallel with this
study.

The present work is motivated by recent experimental observations reported by
Sutter and Lagally (2000) and  Tromp et al. (2000) on the growth of
Si$_{1-x}$Ge$_x$ films on Si(001) substrates  with Ge concentration $x$ in the
range $0.1 \le x \le 0.4$. The experiments have clearly revealed that, during
early stages of growth, shallow stepped mounds whose side walls are made up of
widely spaced steps emerge as an inherent morphological instability of the film.
As more material is deposited, the spacing between the steps gradually decreases
until the sidewalls reach a  certain crystallographic orientation. This type of
nucleationless growth of epitaxial islands is not predicted by a widely used
nucleation barrier
model for the energetics of faceted islands (Tersoff and LeGoues 1994), which
is based on the competition between gain in elastic energy and the energy spent
in creating the surface of the island.  Approaches to surface instability that
account for neither orientation dependence nor strain dependance of surface
energy  have been unable to
explain the observed growth mode. A similar growth mode has also been observed in very
recent experiments on  Ge films grown on Si(001) substrates (Vailionis et al.
2000, Rastelli et al. 2001), where small islands with widely spaced steps
called "pre-pyramids" are the three dimensional features that first appear during
growth; these pre-pyramids eventually evolve to faceted islands.   In this paper,
we will show that the emergence of stepped islands, without a nucleation
barrier, can be explained by incorporating the physics of steps, in particular
their interactions and the dependance of their formation energies on the
mismatch strain.

Before we proceed with development of the continuum description, we  recall a
few results regarding stepped crystal surfaces that are well-established in
surface science.   Below the characteristic roughening temperature, which is in
excess of 1000 $^\circ$C for most semiconductors, a nominally flat surface of a
crystal that is misoriented by a small angle from a high-symmetry direction
consists of a train of straight parallel steps, as illustrated schematically in
\fig{vicinalsurf}. The surface energy  of such a  ``vicinal" surface is given by
\be
\gamma(\theta) = \gamma_0\cos\theta + \beta_1 |\sin\theta| +  \beta_3\frac{|\sin\theta|^3}{\cos^2\theta},
\lab{Vicinalse}
\ee
where $\theta$ is the misorientation angle, $\gamma_0$ the energy density of the
atomically flat surface and the parameters $\beta_1$ and $\beta_3$ are
related to step creation and interaction energies respectively.
Specifically, $\beta_1 = \beta/h_s$, where $\beta$ is the energy to
create a unit length of an isolated step and $h_s$ is the height of an
atomic step. Like any other defect, say a dislocation or a
vacancy, steps give rise to long-range stress fields in the solid;
the steps interact
with each other, as well as with other defects in the material,
through these fields. The amplitude of the stress
field produced by a step decays with the inverse square
of the distance from the step (Marchenko and Parshin 1980). The
last term on the right in \eq{Vicinalse} arises from the interactions between
steps through their stress fields.

The discussion proceeds in the following way. In Sec.~2 we will derive the
energy of a stepped surface for small strains but arbitrary orientations,
starting from the surface energy of a vicinal surface given in \eq{Vicinalse}.
To facilitate calculations, this energy  will then be simplified so that it
applies to surfaces whose deviation from a flat orientation is small. The
surface chemical potential which represents the driving force for change in
shape due to  mass transport on the surface will be derived in Sec.~3. In
Sec.~4, we develop a variational formulation of surface evolution and apply it
to study the growth of a thin film on a lattice mismatched substrate under the
influence of a constant deposition flux. Our focus will be on the evolution and
interactions of stepped mounds that appear on the substrate surface as soon the
deposition flux is turned on. It will be shown that the spacing between the
steps on the sidewalls of these mounds decreases continuously until it reaches
an optimum value, determined by the competition between repulsive step
interactions and a strain induced lowering of step formation energies.  We
provide a summary of the key results along with future directions of research in
Sec.~5. The connections between the continuum description developed in this
paper and discrete elastic models of steps will be discussed in Appendix A. In
Appendix B, we calculate the energy of a strained island using both the
continuum and discrete descriptions of stepped surfaces. Here, we show that  the
mismatch strain can lead to lowering of the surface energy of the vicinal
surfaces and can lead to elimination of the barrier to nucleation of these
islands.

\section{Energy of a strained solid with a stepped surface}

Consider a strained crystal that occupies the region $R$ and that undergoes
two-dimensional generalized plane strain deformation.  The region is bounded by
a free surface $S$ as shown in \fig{csteps}; the remainder of the boundary of
$R$ is ``workless"  due to symmetry constraints or its remoteness. The free
surface is viewed as a being made  up of an array of infinitely long, straight
monatomic steps of height $h_s$. Initially, the surface $S$ is  atomically flat
with its normal vector in the $x_2-$direction. The goal of this section is to
represent the change in energy associated with departure from flatness in terms
of the evolving surface shape.

When the surface $S$ is flat, the crystal is subjected to a spatially uniform
strain $\epsilon^0_{ij}$.
To ensure that $S$ is stress-free in this configuration, this initial
strain must satisfy
\be
C_{i2kl}\epsilon^0_{kl} =0,
\ee
where $C_{ijkl}$ is the array of elastic constants.
The total strain in the crystal is then written as
$\epsilon_{ij} = \frac{1}{2}(u_{i,j}+ u_{j,i}) + \epsilon^0_{ij}$, so that
$u_i$ represents the {\it additional} displacement field that arises as the surface deviates from
its flat orientation. The material stretch along the evolving surface can
be expressed in terms of the limiting value of the bulk strain on the surface as
$\epsilon = m_i\epsilon_{ij}m_j$, where $m_i$  is a
unit vector  that is locally tangent to $S$. For the surface of the stressed
solid under
consideration, as illustrated schematically in \fig{csteps}, we can generalize \eq{Vicinalse} to include the effects of
surface stretch $\epsilon $ to express the surface energy density as
\be
\gamma[h(s),\epsilon(s)] = \left(\gamma_0 + \tau_0 \epsilon(s)\right)\sqrt{1-h^{\prime}(s)^2} +
 \left(\beta_1 +  \tilde{\beta_1}\epsilon(s)\right)\left|h^{\prime}(s)\right| +
 \beta_3\frac{\left|h^{\prime}(s)\right|^3}{1-h^{\prime}(s)^2},
\lab{Continuumse}
\ee
where $h(s)$ is the deviation of the height of the surface from its flat
orientation as a function of arclength $s$ along $S$, $\tau_0 = \gamma_0 +
d\gamma_0/d\epsilon$ is the
surface stress of the flat surface at the current level of stretch, and
$\tilde{\beta_1} = d\beta_1/d\epsilon$ is a measure of the sensitivity of the
formation energy of a step on a strained surface.  The last term has the
same interpretation as in \eq{Vicinalse} and its inclusion is essential.  In principle, we should
also expand the coefficient of  $\left|h^{\prime}(s)\right|^3$ in terms of
surface stretch. However, it will be shown in the appendix, on the basis of direct comparison with discrete
step models, that ignoring the  higher order terms in the expansion provides
a reasonable approximation. Consequently, only the
leading order term represented by  $\beta_3$  is retained.

\begin{figure}[h]
\begin{center}
\includegraphics[width=11cm, height=11.0cm]{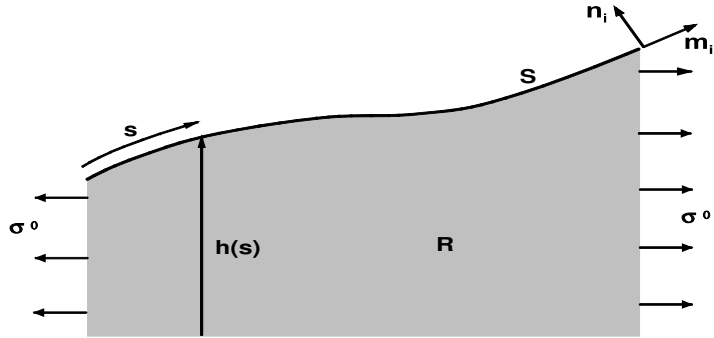}
\vspace{-5.0cm}
\parbox{1.0cm}{\ }\
\parbox{14cm}{
\caption{\lab{csteps}Configuration of the stepped surface $S$ of a strained crystal $R$. The deviation of the height of the surface from
its initially flat orientation is denoted by $h(s)$ where $s$ is the
arclength of the surface along $S$. }
}
\end{center}
\end{figure}

The total energy of the solid is the sum of the elastic energy
and the surface energy,
\be
E[h(s),\epsilon_{ij}({\bf{x}})] = \frac{1}{2}\int_RC_{ijkl}\epsilon_{ij}({\bf{x}})\epsilon_{kl}({\bf{x}})d{\bf{x}} + \int_S \Sigma(s) ds + \int_S \tau(s) \epsilon(s)ds,
\lab{Continuumfe}
\ee
where $\bf{x}$ is any material point in $R$, and $\Sigma(s)$ and $\tau(s)$
are the local surface energy and surface stress at zero
surface stretch.  The latter quantities
are given by
\be
 \Sigma(s) = \gamma_0 \sqrt{1-h^{\prime}(s)^2}+ \beta_1\left|h^{\prime}(s)\right| + \beta_3\frac{\left|h^{\prime}(s)\right|^3}{1-h^{\prime}(s)^2}
\ee
 and
\be
\tau(s) = \tau_0 \sqrt{1-h^{\prime}(s)^2}+ \tilde{\beta_1}\left|h^{\prime}(s)\right|,
\ee
respectively.  Because mechanical equilibrium  is achieved very quickly
on the time scale associated with diffusive
mass transport on crystal surfaces, we will first derive conditions for
equilibrium at fixed surface shape $h(s)$.
Then, by using the equilibrium fields obtained,
we can proceed to derive an expression for the change in the energy of
the system in terms of surface shape change, which leads
naturally to a corresponding definition of surface chemical potential
as an energetic driving force for surface shape change.

Mechanical equilibrium requires that the
variation in the free energy $E$ at fixed surface shape $h(s)$ due to a small
perturbation in displacement field $\delta u_i$ from equilibrium values of $u_i$
must  vanish to linear order in
$\delta u_i$. The first order variation in energy in the present case is
\be
\delta E =  - \int_RC_{ijkl}u_{l,kj}\delta u_i d{\bf{x}} +\int_SC_{ijkl}\left(u_{l,k} + \epsilon^0_{kl}\right)n_j\delta u_i ds + \int_S
\tau(s)m_i\delta u_{i,j} m_j ds,
\ee
where $n_i$ is the local outward unit normal vector to $S$.  The last term can be conveniently rewritten as
\be
\int_S
\tau(s)m_i\delta u_{i,j} m_j ds = \int_S
\tau(s)m_i\frac{d\delta u_{i}}{ds} ds.
\ee
Because the bounding surface of $R$ is workless, we can rewrite this
expression in the form
\be
\int_S
\tau(s)m_i\frac{d\delta u_{i}}{ds} ds = -\int_S
\frac{d(\tau(s) m_i)}{ds}\delta u_{i} ds.
\ee
Noting that
\be
\frac{d(\tau(s) m_i)}{ds} = \frac{d\tau(s)}{ds}m_i + \tau(s)\kappa(s)n_i,
\ee
where $\kappa(s)$ is the curvature of the surface, which is taken as positive if the
center of curvature is outside the material, we can rewrite the variation in the total energy as
\be
\delta E = - \int_RC_{ijkl}u_{l,kj}\delta u_i d{\bf{x}} + \int_S\left[C_{ijkl}\left(u_{l,k} + \epsilon^0_{kl}\right)n_j - \frac{d\tau(s)}{ds}m_i -
\tau(s)\kappa(s)n_i\right]\delta u_i ds .
\ee
Because the variation must vanish for arbitrary $\delta u_i$, we obtain  the
usual equilibrium
condition that the divergence of stress must vanish at each material point,
$C_{ijkl}u_{l,kj} = 0$ in the region $R$.  In addition, traction boundary
conditions reflecting the structure of the surface emerge in the form
\begin{eqnarray}
C_{ijkl}u_{l,k}n_jm_i &=& -C_{ijkl}\epsilon^0_{kl}n_jm_i + \tilde{\beta_1}h^{\prime \prime}(s){\mathrm{ Sgn}}[h^{\prime}(s)] -\frac{\tau_0h^{\prime}(s)h^{\prime\prime}(s)}{\sqrt{1-h^{\prime}(s)^2}}, \nonumber \\
C_{ijkl}u_{l,k}n_jn_i &=& -C_{ijkl}\epsilon^0_{kl}n_jn_i +
\left(\tau_0 \sqrt{1-h^{\prime}(s)^2}+ \tilde{\beta_1}\left|h^{\prime}(s)\right|\right)\kappa(s),
\lab{Strac}
\end{eqnarray}
where ${\mathrm{Sgn}}[p] =
d|p|/dp$  for any quantity $p\neq 0$.  These must be satisfied on $S$ by
the elastic field.  The first terms
on the right sides of \eq{Strac} represent,
respectively, the normal and shear tractions applied to the surface to compensate for the
relaxation of the bulk
stress $C_{ijkl}\epsilon^0_{ij}$ due to surface reorientation; these tractions
ensure that the surface $S$
remains traction free. The other terms on the right side of (\ref{Strac})$_1$ represent the shear traction
due to variation of
surface stress with position along the surface, whereas the second term in
(\ref{Strac})$_2$ represents the normal traction on $S$ due to the so-called
Laplace pressure  which arises from surface stress. Note that these conditions
arise only as a result of the assumption that energies of surface features
may depend on strain.  They involve no other ad hoc assumptions.

Up to this point, there has been no restriction on the magnitude of the surface
shape change.  From this point onward, however, attention will be restricted
to surface shapes that are locally misoriented by only a small angle from the
flat orientation, that is,  the condition $|h^{\prime}(s)| \ll 1$ prevails everywhere on $S$.
Using the
fact that $m_{1}=1-h'(x_{1} )^{2}/2 =n_{2}$ and $m_{2}=h'(x_{1})=-n_{1}$
to second-order in the slope, the boundary conditions become
\begin{eqnarray}
C_{12kl}u_{l,k} &=& C_{11kl}\epsilon^0_{kl}h'(x_1) + \tilde{\beta_1}h''(x_1){\mathrm{ Sgn}}\left[h'(x_1)\right] \equiv f_1(x_1), \nonumber \\
C_{22kl}u_{l,k} &=&  \tau_0h''(x_1)\equiv f_2(x_1),
\lab{Surfacetrac}
\end{eqnarray}
where $f_1$ and $f_2$ are the components of surface traction in the coordinate
directions that give rise to stress fields in the crystal  associated with shape change. The displacement
field at any point in the crystal can now be obtained using the elastic half-space Green's function $G_{ij}(\bx)$ as
\be
u_i(\bx) = \int_{-\infty}^{\infty}G_{ij}(\bx - x^{\prime}\bE_1)f_j(x^{\prime})
\,dx^{\prime}.
\lab{Contdisp}
\ee

We can now recast the minimum free energy at fixed surface shape in terms of the surface tractions introduced in \eq{Surfacetrac}.
Applying the divergence theorem to the first term in \eq{Continuumfe},
recalling the workless nature of the boundary constraints, and using the fact that
\begin{eqnarray}
\left(\tau_0+\tilde{\beta}_1\left|h'\right|\right)m_i\epsilon^0_{ij}m_j &=& \left(\tau_0 +
\tilde{\beta}_1\left|h'\right|\right)\epsilon_{11}^0
+ 2\tau_0\epsilon_{12}^0h'
\nonumber \\
&& + \left(\tau_0(\epsilon_{22}^0-\epsilon_{11}^0) + 2\tilde{\beta}_1\epsilon_{12}^0\mathrm{Sgn}\left[h'\right]\right)\left(h'\right)^2 + O(h^3)
\end{eqnarray}
we find that
\begin{eqnarray}
E[h(x)] &=&
E_0+ \int_{S_1}\left(\beta_1 + \tilde{\beta_1}\epsilon_{11}^0 + 2\tau_0\epsilon_{12}^0\mathrm{Sgn}\left[h'(x)\right]\right)\left|h'(x)\right|dx + \int_{S_1}\beta_3 \left|h'(x)\right|^3 dx \nonumber \\
& +& \int_{S_1} \left(\tau_0(\epsilon_{22}^0-\epsilon_{11}^0) + 2\tilde{\beta_1}\epsilon_{12}^0\mathrm{Sgn}\left[h'(x)\right]\right)\left(h'(x)\right)^2 dx
-\frac{1}{2}\int_{S_1}f_i(x)u_i(x)dx,
\lab{Continuume}
\end{eqnarray}
where $E_0$ is the total energy when the surface is flat, $S_1$  refers to the projection of $S$
in the $x_1$-direction and the
coordinate $x_{1}$ has been replaced by $x$ for simplicity.  This is
the total free energy of the system with a perturbed surface shape.

Before proceeding to study the evolution of stepped surfaces,
we make a few observations about the continuum representation of system
free energy change given by \eq{Continuumfe} and \eq{Continuume}:
\begin{enumerate}

\item All the terms in \eq{Continuume} except the last are local, in the sense
that the energy density at any point  depends only on the local surface slope.
The last term, however, represents the interaction of the surface force
distributions given in \eq{Surfacetrac}; in the jargon of surface science, these
represent force monopoles. Since these monopole interactions are long-ranged,
the energy density  at a point on the surface depends on the surface shape at
all points on the surface.

\item The free energy in the small slope approximation is correct to $O(h^2)$;
only one term of $O(h^3)$ has been retained.  It will be shown in the Appendix A
that this approximation where all the other $O(h^3)$ terms are ignored is a
reasonable one, by means of comparison of the continuum description with
discrete step models.

\item The small slope approximation is not essential to the development. If this
approximation is not made, the total energy can be  evaluated by solving the
boundary value problem, which consists of the bulk equilibrium equations
subject to boundary conditions  \eq{Strac}, using a numerical technique such as
the finite element or boundary element method.

\item The quantity $\tilde{\beta_1}$ must be either positive or negative, if the
step formation energy is strain sensitive.  It follows that
$\beta_1+\tilde{\beta_1}\epsilon^0_{11} < 0$  for compressive or tensile strain
$\epsilon^0_{11}$, respectively.  This implies that the energy of the stepped
surface becomes {\it lower} than the energy of the surface  without steps for
strain of one sign or the other. A plot of the surface energy of a vicinal
surface for which  $\epsilon_{12}^0 =0$ is shown in \fig{anisose}. In this case,
when strains are  sufficiently compressive with $\tilde{\beta_1} > 0$, the
variation of surface energy with surface orientation $\theta$, as given by \be
\gamma(\theta) = \gamma_0 + \tau_0\epsilon_{11}^0  + \left(\beta_1 +
\tilde{\beta_1}\epsilon_{11}^0\right)|\theta| +
\tau_0(\epsilon_{22}^0-\epsilon_{11}^0)\theta^2 + \beta_3 |\theta|^3,
\label{strainse} \ee  develops minima away from $\theta=0$. At small
misorientation angles, the  step formation energy dominates the repulsive
interaction energy between the  steps, while at larger angles, the step
interactions are larger in magnitude.  The competition between these two
opposing effects results in optimum  misorientation angles denoted by
$\theta=\pm \theta^*$ in \fig{anisose}.  It follows that $\tilde{\beta_1}$ is a
key quantity in determining the morphology  of strained crystals\footnote {The
value of $\tilde{\beta_1}$ cannot be deduced from a continuum  analysis.
However, by means of atomistic simulations of stepped surfaces on strained
crystals of Si being conducted in parallel with this study, we have concluded
that the anticipated minima do indeed arise within the strain  range of
interest.}. Interestingly, since the force monopole in \eq{Surfacetrac} has a
contribution  that depends on $\tilde{\beta_1}$, this key quantity also enters
the non-local part of the total energy.

\item If the energy of a stepped surface under strain becomes lower than the
energy of a flat surface, there is no barrier to nucleation of  epitaxial
islands whose sidewalls are made up of surfaces with widely spaced steps.  In
Appendix B, we consider the energetics of such stepped mounds  using both the
discrete models of steps and the present continuum description.

\begin{figure}[h]
\begin{center}
\includegraphics[width=9cm,height=9cm]{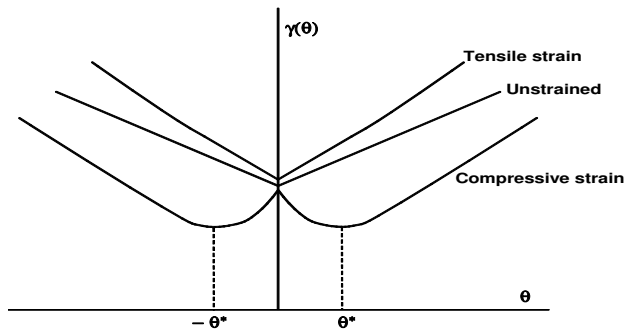}
\vspace{-4.0cm}
\parbox{1.0cm}{\ }\
\parbox{14cm}{
\caption{\label{anisose}Dependence of the surface energy $\gamma(\theta)$ of a vicinal surface
on surface orientation $\theta $
for $\tilde{\beta_1} > 0$, $\tau_0 > 0$ and $\epsilon_{12}^0=0$. When
compressive strains are sufficiently large, the energy of a stepped surface
becomes lower than the energy of a flat surface. In this case, competition
between strain-induced lowering of the formation energy of the steps and the
repulsive step interactions results in a minimum in the surface energy at an
optimum angle, $\theta^*$ indicated in the figure.}}
\end{center}
\end{figure}

\end{enumerate}

\section{Surface chemical potential}

The evolution of surface morphology of strained crystals as a result of mass
transport by surface  diffusion or other mechanism has been studied extensively
on the basis of continuum models,  beginning with the introduction of a surface
chemical potential by Herring (1953).  The essential  ingredients for
describing evolution are the chemical potential field defined as a function of
position  over the evolving surface of the solid and a kinetic relationship
between surface mass flux and  the gradient in chemical potential.  The rate of
change of surface shape then follows from conservation of mass.  The kinetic
relationship must be defined in such a way as to  ensure that the free energy
decreases as surface evolution progresses.  In this section, the  surface
chemical potential for a stepped surface will be obtained from the free energy
expression  given in (\ref{Continuumfe}).  The result is written first for
arbitrary magnitude of surface slope, and it is then  reduced to a simpler form
applicable for the case of surface perturbations with slopes of  small
magnitude.

In deriving the mechanical equilibrium equations in Section 2, it was assumed
that the surface  was fixed with respect to the material.  This was justified by
the observation that mechanical fields  equilibrate very fast on the timescale
of surface evolution.  We now assume that the surface  shape depends on time
$t$, and we consider the associated rate of change the free energy  of the
system as the
surface  shape changes, with the system being continuously in mechanical
equilibrium.  The rate of change of this free energy measure then
has the form
\be
\dot{E}(t)  = \int_S \mu[h(s,t)] v_n(s,t)ds,
\lab{Chempot}
\ee
where $v_n(s,t)$ is the outward normal velocity of the surface $S$ with
respect to the material instantaneously on it and
$\mu[h(s,t)]$ is the local chemical potential.  At any point along the surface,
the quantity $v_n\,ds$ is the rate of addition of material volume to the surface.
It follows that its coefficient is the change in free energy per unit volume
of material added, that is, the coefficient $\mu$ represents the surface chemical potential field
(Freund 1998).

We can write the rate of change of free energy (\ref{Continuumfe}), following application of the
divergence theorem to the first term, as
\be
\dot{E} =  \int_S \left(C_{ijkl}\epsilon_{kl}\dot{u}_i n_j+ U v_n \right) ds + \int_S \left(
\frac{\partial\Sigma}{\partial\theta}\frac{d\theta}{dt} + \frac{\partial\tau}{\partial\theta}\frac{d\theta}{dt} \epsilon + \tau \frac{d{\epsilon}}{dt}\right)ds
-\int_S\left(\Sigma + \tau \epsilon\right)\kappa v_n ds,
\lab{Fechange}
\ee
where $U =\frac{1}{2}C_{ijkl}\epsilon_{ij}\epsilon_{kl}$ is the elastic
strain energy density, a superposed dot denotes a material time
derivative, and $d\ /dt$ denotes the total time derivative whenever it must be distinguished from the
material derivative.  For example, the rate of change of surface stretch
\be
\frac{d{\epsilon}}{dt} = \dot{\epsilon} + m_im_j\epsilon_{ij,k}n_kv_n
\ee
includes a convective
term,
and the rate of change of surface orientation following the non-material
surface requires that
\be
{d\theta\over dt} = {\partial v_n \over \partial s}\quad \mbox{and}
\quad {d m_i \over dt} = n_i {\partial v_n \over \partial s}.
\ee
The first term in (\ref{Fechange}) accounts for the change in strain energy in the crystal due to surface
evolution, whereas the remaining terms account for the change in surface energy.  Recalling that
the system is enclosed within workless boundaries, we can rewrite (\ref{Fechange}) in the form (\ref{Chempot}) with
\be
\mu[h(s,t)] = U -\kappa\left[ \Sigma + \epsilon  \tau + \frac{\partial^2\Sigma}{\partial\theta^2} + \epsilon \frac{\partial^2 \tau}{\partial\theta^2}\right]-\frac{\partial \tau}{\partial \theta}\frac{\partial \epsilon}{\partial s} - 2\frac{\partial}{\partial s}(m_in_j\epsilon_{ij}\tau) + m_im_j\epsilon_{ij,k}n_k\tau.
\ee
In the limit of small slope $| h,_x(x,t)|\ll 1$, where the subscript $x$ following the comma indicates
partial differentiation with respect to $x$, this expression can be further simplified to the form
\begin{eqnarray}
\mu[h(x,t)] &=& U_0
 -h,_{xx}\left[ (\beta_1 + \tilde{\beta}_1\epsilon^0_{11})\delta\left(h,_x \right) +2\tau_0(\epsilon_{22}^0 - \epsilon_{11}^0) + 4 \tilde{\beta}_1\epsilon_{12}^0\mathrm{Sgn}\left[h,_x\right]  + 6\beta_3\left| h,_x \right|\right] \nonumber \\
&& + C_{11kl}\epsilon^0_{kl}u_{1,x} -\tilde{\beta}_1  \mathrm{Sgn}\left[h,_x\right]u_{1,xx} - \tau_0u_{2,xx},
\lab{SSchemp}
\end{eqnarray}
where $U_0$ is the value of $U$ when the surface is flat and $\delta (\ )$ is the Dirac delta function.

A vexing feature of the chemical potential function obtained in (\ref{SSchemp}) is the appearance of the Dirac
delta function in the term representing step creation energy.  We expect this term to play a role in
evolution of the surface of the strained crystal because the step creation energy can become
negative for strain of one sign or the other, if the magnitude is sufficiently large.  While it is
possible to introduce any number of ad hoc ``regularizations" to deal with this singularity, it is not
evident that the results obtained will be independent of the particular choice made.  Therefore, we
adopt a variational framework for describing surface evolution which is based on the chemical
potential derived but which circumvents the problem posed by the singular step creation energy
term.

\section{A variational approach for strain-driven  surface evolution}

In this section, we apply the continuum description of the energy
derived in \eq{Continuume} to study  the growth of a strained film
bonded to a lattice-mismatched substrate.  The constraint of the
substrate maintains the mismatch stress in the film. As noted before,
recent experiments (Sutter and Lagally 2000 and Tromp et al. 2000) have observed that  growth of the deposited film
proceeds via formation of shallow mounds whose stepped side-walls
ultimately evolve to a faceted orientation. The evolution takes place
through diffusion of atoms across the terraces that separate the steps
on the side-walls of these mounds.

In the case of terrace diffusion-limited kinetics, the mass flux on
the surface can be related to the
gradient of the chemical potential derived in \eq{SSchemp} through
\be
j(x,t) = - c  \mu_{,x},
\lab{Stranseq}
\ee
where $c $ is a coefficient representing surface mobility of
diffusing species. If there is a constant deposition flux $f$, mass conservation can be invoked to connect the
mass flux to the surface shape through the relation
\be
h_{,t} + j_{,x} = f.
\lab{massc}
\ee

Our goal is to develop a variational
framework for modeling the evolution of these stepped surfaces.  If we
focus attention on surface profiles that are periodic with
wavelength $\lambda$, the thickness of the film $h(x,t)$, which is
illustrated in \fig{film}, can be
expressed as
\be
h(x,t) = \sum_{n = 0}^{N_f} a_n(t)\cos(n k x), \,\,\,\,\,\, 0 \le x \le \lambda,
\lab{height}
\ee
where $k = 2 \pi/\lambda$. The wavelength should be large enough to
accommodate several stepped mounds and other features of interest.  The number of terms $N_f$ included in the series
is finite but otherwise unrestricted. The variational formulation provides first order coupled ordinary differential equations
for the Fourier coefficients $a_n(t)$; these equations will be numerically integrated to obtain the evolving surface shape for
$t>0$ using information on the surface shape at $t=0$.

Using \eq{height} in \eq{massc}, we can express the mass flux in the form
\be
j(x,t) = -\sum_{n =1}^{N_f} \dot{a}_n(t)\frac{\sin(nkx)}{nk},
\lab{jfourier}
\ee
and the mean height of the surface as $a_0(t) = ft$.
While \eq{Stranseq} and \eq{massc} constitute a closed
set of equations, they cannot be directly integrated to obtain the
surface shape because of the singular nature of the chemical potential
in \eq{SSchemp}, as noted in the preceding section. We now turn to the variational approach which
provides well-behaved evolution equations for
the Fourier coefficients in \eq{height} by  using the weak form of  \eq{Stranseq}.

\begin{figure}[h]
\begin{center}
\includegraphics[width=10.0cm,height=10.0cm]{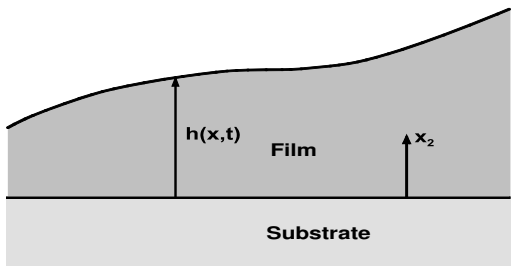}
\vspace{-6cm}
\parbox{1.0cm}{\ }\
\parbox{14cm}{
\caption{\lab{film}Strained film on a relatively thick lattice-mismatched substrate.
The time dependent thickness of the film is denoted by $h(x,t)$.
When the film surface is flat, there is a spatially uniform elastic mismatch stress
in the film and the substrate is stress free. }}
\end{center}
\end{figure}

The variational approach can be formulated (Suo 1997) by noting that the functional
\be
\Phi[j(x,t)] = \int_0^{\lambda} \mu,_x j(x,t)dx + \int_{0}^{\lambda}\frac{j^2(x,t)}{2 c }dx
\lab{Varfunct}
\ee
attains a minimum value when $j(x,t)$ satisfies the kinetic relation \eq{Stranseq}.
The idea is to express the functional
in terms of the time derivatives of the Fourier coefficients $\dot{a}_n(t)$ and
to then minimize the functional with respect to variations in these rates.  The first of these tasks can be accomplished by integrating the first term
by parts and by using \eq{massc} and \eq{jfourier} to obtain
\be
\Phi[\dot{a}_1,\dot{a}_2 \cdots]  = \sum_{n =1}^{N_f}\frac{\partial E[a_0,a_1 \cdots]}{\partial a_n}\dot{a}_n +  \sum_{n =1}^{N_f}\frac{\pi}{2c  n k^3}\dot{a}_n^2,
\label{functional}
\ee
where the total free energy $E[a_0,a_1 \cdots]$ of the film-substrate system
can be directly evaluated from \eq{Continuume} with one minor modification.
Since the mismatch strain is present only in the film but not in the
substrate (refer to \fig{film}),
$\epsilon_{ij}^0$ in \eq{Continuume} has to replaced by a spatially
dependent function $\epsilon_{ij}^0(x_2) = \epsilon_{ij}^0 H(x_2)$,
where $H(x_2)$ is the Heaviside function which has the value of zero when
$x_2 < 0 $ and the value one when $x_2 > 0 $.
Minimizing the functional in \eq{functional} with respect to the $\dot{a}_n$'s,
we get the evolution equations
\be
\dot{a}_n(t) = - \frac{c  n k^3}{\pi} \frac{\partial E}{\partial a_n},
\label{evolutioneq}
\ee
which are unambiguous and well-behaved. We now proceed to integrate these rate
equations to determine the time evolution of the
deposited film. At each time step of integration, the quantity
${\partial E}/{\partial a_n}$ is calculated numerically.
The important role of the nonlinear contribution to (\ref{evolutioneq}) for the convergence of this approach
is illustrated in Appendix C by considering only a single Fourier mode.

\begin{figure}[p]
\begin{center}
\includegraphics[width=14.0cm]{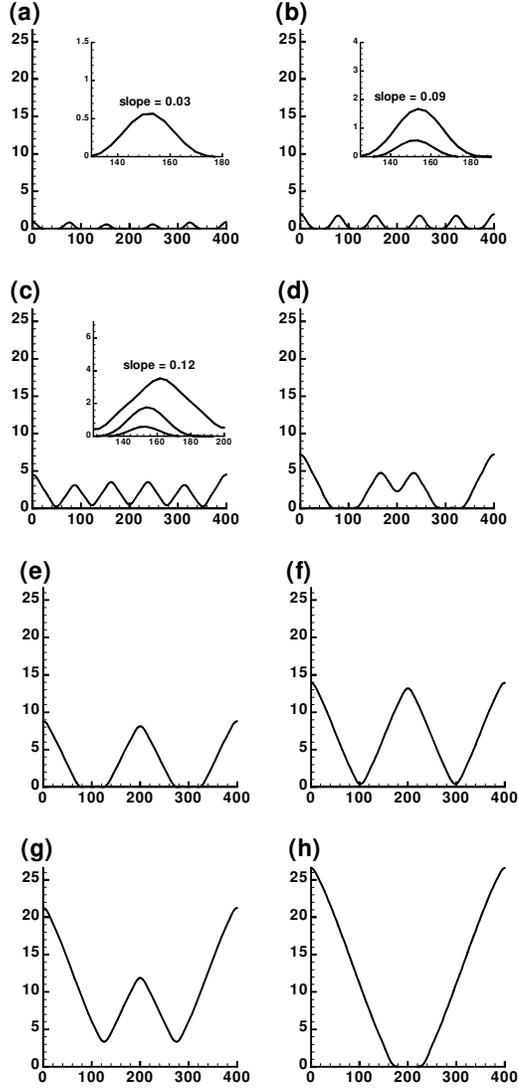}
\vspace{-2.5cm}
\parbox{1.0cm}{\ }\
\parbox{14cm}{
\caption{\label{filmgrowth}The figure shows a time sequence of surface profiles
 of $h(x,t)$ plotted on the vertical axes versus $x$ plotted on the horizontal
 axes  of  a strained film during constant flux growth on a lattice-mismatched
 substrate.  All the dimensions in nanometers. The inserts in (a)-(c) show the
 evolution  of the third island from the right. To aid in the comparison of
 shapes at different times, we have included the island shape from (a) in (b).
 Similarly, we have included the island shapes from (a) and (b) in (c). The
 slope of the largest island  in each of the smaller insets is indicated in the
 figure.}  }
\end{center}
\end{figure}

 The numerical integration of \eq{evolutioneq} was carried out using the
 fourth-order Runge-Kutta procedure with adaptive step-size control as described in detail
  by Press et al. (1992).
The parameters in the continuum description of the energy  were chosen as
follows: $\beta_1 = 0.03\, J/m^2$, $\tilde{\beta_1} = 15\, J/m^2$,
$\tau_0 = 1 J/m^2$, $\epsilon_{11}^0 = -0.01$ and $\beta_3 = 2.86\, J/m^2$.
It is assumed that the substrate and the film are isotropic with similar elastic
properties, and that the Young's modulus and the Poisson's ratio are
$10^{11}\, N/m^2$ and $0.3$, respectively.  Using these parameters, we find
that $\epsilon_{22}^0 = 0.0023$ and $\epsilon_{12}^0 = 0$. Also, for the
compressively strained film, the
surface energy of the film (sketched schematically in \fig{anisose}) attains
a minimum when $\theta^* = 0.12$, which implies that the sidewalls of the stepped mounds would eventually evolve to this angle.

The initial profile of the film was chosen to be a sinusoid of wavelength
$\lambda = 400 $nm, with an amplitude of $0.4$ nm, so that the only
non-vanishing Fourier components at $t = 0 $ are $a_0$ and $a_1$. In our
calculations, sixteen Fourier coefficients were used to keep track of the
surface shape. The evolution of the deposited film is  shown in
\fig{filmgrowth}. As the deposition flux is turned on, the material on the
surface very quickly  gathers into five stepped mounds with slopes much smaller
than the optimum value of $0.12$ as shown in \fig{filmgrowth}(a). The insert in
\fig{filmgrowth}(a-c)  tracks the evolution of one of these mounds. As more
material is deposited, the side-walls of the mound become steeper until they
reach the optimum angle $\theta = 0.12$ in \fig{filmgrowth}(c). It can also be
observed that the center of mass of the mound shifts gradually to the right in going from
\fig{filmgrowth}(a) to \fig{filmgrowth}(c). This can be understood by looking
at the interactions between the islands. Since elastic relaxation is achieved
for widely spaced islands, the islands tends to repel each
other (Floro et al. 2000). It can be seen that the island in \fig{filmgrowth}(a) is located closer
to the island on its left and would therefore tend to shift towards the right
via diffusion of atoms from the side-wall on the left to the one on the right.
Once the side-walls of the islands reach the optimum orientation, they grow in
a self-similar fashion, until they come in contact  with their neighbors as
shown in \fig{filmgrowth}(c). At this point, self-similar coarsening is
initiated, which leads to a decrease in areal coverage (fraction of the surface
of the substrate covered by the film) as is evident in \fig{filmgrowth}(e).
Here there are two islands, with side walls at $\theta = 0.12$, separated by
50nm. As more material is deposited, these further grow in size and,  eventually,
one of the islands grows at the expense of the other island as shown in
\fig{filmgrowth}(f)-(h).

The evolution of the film in the early stages of island growth are in close
agreement with the recent experiments of Sutter and Lagally (2000) and Tromp et
al. (2000). The key result of these experiments is that the islands evolve as a
{\it natural instability} without any nucleation barrier. As we show in the
Appendix, there is no barrier for nucleating stepped islands if the strains are
compressive. Islands with orientations below the optimum facet orientation
lower the energy of the film and provide a kinetic pathway for obtaining
faceted islands that is free of any nucleation barrier. This is indeed what is
seen in experiments and during early stages of growth in
\fig{filmgrowth}(a-c).

It can be seen from Appendix~B that the energy of the islands is dominated by
the surface energy at small island volumes and by the elastic energy at large
volumes. Using the parameters adopted in our calculations, we can use
\eq{eisland} to show that  the crossover between these two regimes takes place
when the base width of the island is about 200-300nm. When the island sizes are
smaller than this value, the sidewalls are oriented at the optimum angle that
minimizes the surface energy. With increasing base width, it is known that the
islands undergo shape transitions (Medeiros-Ribeiro et al. 1998), where the sidewall angles change to a
steeper orientation, which is usually a low-energy crystallographic orientation
that makes a larger angle with the substrate. Since such orientations are not
yet included in our model, we do not observe such transitions. We do, however,
find that the slope of the sidewalls of the large islands in
\fig{filmgrowth}(e) is about $0.15$, which is about 25\% bigger than the
optimum slope. This indicates that elastic energy of these larger islands is
becoming comparable to the surface energy.

There is no fundamental impediment
to including additional low energy surface orientations in order to study
steeper sidewall facets.  For example, these might appear as additional
relative minima in the variation of $\gamma(\theta)$ with orientation $\theta$
at angles greater in magnitude than $\theta^*$.  This will be subject of
future investigations.

\section{Summary}

 In summary, we have developed a continuum framework to model the energetics and
 evolution of stepped surfaces of nanostructures. The surface energy of strained
 surfaces is obtained by a generalization of the energy of a vicinal surface to
 account for the effects of stretching of the surface caused by the  mismatch
 strain. We find that the dependence of the formation energy of steps on the
 sign of mismatch strain plays a crucial role in establishing the morphology
 of evolving nanostructures. While this dependence has been seen in
 atomic-scale simulations, we have developed a consistent continuum framework
 that naturally accounts for this crucial effect in modeling  surface evolution.
 Furthermore, we have shown in Appendix~A that our continuum description
 provides a direct way to analyze stepped surfaces, without evaluating complicated
 sums involving discrete steps.

   The continuum framework was applied to study the evolution of an epitaxial
   film bonded to a lattice-mismatched substrate. Our simulations show that the
   deposited material initially gathers in shallow stepped mounds whose
   side-wall angles eventually evolve to an  orientation that is determined by the
   competition between strain-induced lowering of the step formation energy and
   the repulsive interactions between the steps. This kinetic pathway has no
   nucleation barrier and occurs as a natural instability, in agreement with the
   recent experimental observations. The simulations were also used to look at
   self-similar growth of islands leading to their impingement with neighboring
   islands and subsequent coarsening.

 In this article, discussions were limited to two-dimensional deformation fields
 and evolution of surfaces in one dimension through the diffusion atoms, all of
 the same type. Work is currently in progress to extend the current formulation
 to study the evolution of two-dimensional surfaces. Also, since the films of
 interest in device applications are typically alloys of two different
 semiconductors, further insight into surface evolution can be obtained by
 including the possibility of alloy segregation in binary materials.

\section*{Acknowledgments}

{\small \setlength{\baselineskip}{11pt}{The research support of the National
Science Foundation through grant CMS-0093714 and the Brown University MRSEC
Program, under award DMR-0079964, is gratefully acknowledged. }}
\bigskip
\appendix

\noindent{\bf {\Large Appendices}}

\section{Discrete surface step models}

 In developing a continuum description of stepped surfaces, we started with the
 surface energy of a strained vicinal surface given by  \eq{Continuumse}, which
 is a generalization  of the surface energy of the vicinal surface to include
 the effects of surface stretch. The key idea of this section is to demonstrate
 that if one chooses to start with elastic models of discrete steps and
 proceeds to construct a {\it homogenized} continuum description, then the
 elastic displacement fields (given in \eq{Contdisp}) and the  energy of
 stepped surfaces (given in \eq{Continuume}) are recovered. This procedure is
 very instructive, as it will allow us to give physical interpretation for
 parameters such as $\tilde{\beta}_1$ and $\beta_3$ in the continuum description
 in terms of quantities that determine the elastic fields of discrete
 atomic steps. As we will see, these quantities depend on bonding of the atoms
 near the step edges and therefore have to be determined using an atomic level
 calculation or, alternatively, from experiments.
In what follows, we will first
 look at the elastic displacement fields  and the corresponding strain energy
 stored in the crystal with a traction-free stepped surface. As noted earlier,
 the contribution of the strain energy to the total free energy enters through
 the non-local contribution in \eq{Continuume}. We will then proceed to analyze
 the local part of the free energy that depends on the local slope of the
 surface.

\subsection{Non-local parts of the free energy and elastic displacement fields}

 In their pioneering work on step interactions, Marchenko and Parshin (1980) analyzed the
 strain field of a surface step by viewing the step
as a point dipole on the surface of strength ${\bf D} = D_1\bE_1 + D_2\bE_2$,
with components in both the coordinate directions perpendicular to the step
which lies along the $\bE_3$-direction in the present case. Since the $\bE_3$ component of the
moment should vanish in equilibrium, they concluded that $D_2 = \pm \tau_0 h_s$,
where $\tau_o $ is the surface stress and $+(-)$ denotes a positive(negative)
step\footnote{For a positive step located at $x_0$, the surface height satisfies
$h(x_0^+) - h(x_0^-) = h_s$, where $h(x_0^+)$ ($h(x_0^-)$) is the surface
height slightly to the right (left) of the step, while for a negative step the
relation $h(x_0^+) - h(x_0^-) = -h_s$ is satisfied.}
respectively. This value of $D_2$ ensures that the moment created by the
surface stress is compensated. The other component, $D_1$,
depends on details of atomic bonding around the step edge and has to
be obtained from an atomic level calculation or
from experiments.  The 1-component is a dipole without moment, whereas
2-component is a dipole with moment, as noted. If we focus attention on
crystal structures that are
invariant under
rotations of 180$^\circ$ about the $\bE_2$ axis, it is clear that a
positive
step becomes a negative step under
such rotations. While the 2-component flips its sign under such a rotation,
the 1-component does not. Therefore, on
such high symmetry surfaces  (for example, the (100) surfaces of cubic materials commonly employed in strained epitaxy),  both positive and negative steps have identical values for $D_1$.
In addition to these components of the force dipole, for a strained crystal
equilibrium requirements can be invoked to show that a monopole
 $M =  h_s C_{11kl}\epsilon^0_{kl}$ in the $\bE_1$-direction must be included,
 in addition to the point dipole, to
model the stress fields of the step associated with the mismatch strain (Tersoff et al. 1995).
We can now use the surface Green's function to write the elastic displacement field of an isolated step located at the origin  as
\be
u_i({\bx}) = \pm M G_{i1}(\bx) + D_1\frac{\partial G_{i1}(\bx)}{\partial x_1}
\pm D_2\frac{\partial G_{i2}(\bx)}{\partial x_1}
\ee
for a positive/negative step.

 If we consider
the surface shown in \fig{csteps}, with a continuous distribution of steps, the surface forces can be modeled
using continuum monopole and dipole densities
\be
m(x) = \frac{M}{h_s} \frac{dh}{dx} \,\,\,\,\,\, {\mathrm{and}} \,\,\,\,\, {\bf{d}}(x) = \frac{D_1}{h_s}\left|\frac{dh}{dx}\right|\bE_1 + \frac{D_2}{h_s} \frac{dh}{dx}\bE_2
\ee
 respectively. The displacement fields in the crystal due to of distribution of steps can be written as a superposition of the displacement fields of individual steps, or
\begin{eqnarray}
u_i({\bx}) &=& \int_{-\infty}^{\infty}\left[ \frac{M}{h_S}G_{i1}(\bx-x^{\prime}\bE_1)\frac{dh(x^{\prime})}{dx^{\prime}} - \frac{D_1}{h_s}\frac{\partial G_{i1}(\bx-x^{\prime}\bE_1)}{\partial x^{\prime}}\left|\frac{dh(x^{\prime})}{dx^{\prime}}\right|\right. \nonumber \\
&& \left. - \tau_0\frac{\partial G_{i2}(\bx-x^{\prime}\bE_1)}{\partial x^{\prime}}\frac{dh(x^{\prime})}{dx^{\prime}}\right]dx^{\prime}.
\lab{Discrdisp}
\end{eqnarray}
Once the second and third terms are integrated by parts, it is
easily seen that \eq{Discrdisp} can be written as
\begin{eqnarray}
u_i({\bx}) &=& \int_{-\infty}^{\infty}\left[ \frac{M}{h_S}G_{i1}(\bx-x^{\prime}\bE_1)\frac{dh(x^{\prime})}{dx^{\prime}} + \frac{D_1}{h_s} G_{i1}(\bx-x^{\prime}\bE_1)\frac{d^2h(x^{\prime})}{dx^{\prime2}}{\mathrm {Sgn}}\left[\frac{dh(x^{\prime})}{dx^{\prime}}\right]\right. \nonumber \\
&& \left. + \tau_0G_{i2}(\bx-x^{\prime}\bE_1)\frac{d^2h(x^{\prime})}{dx^{\prime2}}\right]dx^{\prime},
\lab{Discrdisp1}
\end{eqnarray}
which is in the form given in \eq{Contdisp} with the surface traction $\bF$ precisely as given by \eq{Surfacetrac} if we identify
$\tilde{\beta_1}$ with $D_1/h_s$. It can also be verified that the contribution to the free energy from this
displacement field is
\be
E_{non-local} = -\frac{1}{2}\int_{S_1}{\bf f}\cdot{\bf u}\,\, dx_1,
\ee
which is nothing but the last term in \eq{Continuume} which represents the non-local contribution to the free energy.

To summarize, an interesting outcome of the simple homogenization process that we have
carried out is that even though individual steps are modeled using both
monopoles and dipoles, the displacement fields are determined solely in terms
of a distribution of monopole-like surface tractions given in \eq{Surfacetrac}.
The physical reason for this can be understood by appealing to the analogous
homogenization problem of dipoles in electrostatics or magnetostatics. In a
polarized dielectric with dipole moment density ${\bf P}(\bx)$, the homogenized
electric fields can be obtained by replacing the spatially varying dipole
density with an equivalent charge density $\rho(\bx) = -\nabla \cdot {\bf
P}(\bx)$; this means that a spatial variation of the dipole density physically
corresponds to charge accumulation. In the case of steps, the force monopole
distribution that originated from the dipoles in \eq{Surfacetrac} is seen to be
nothing but $d{\bf d}(x)/dx$, where ${\bf d}(x)$ is the dipole density
introduced earlier. It is also clear that a homogenous distribution of dipoles
does not give rise to any long-range elastic displacement fields; if we consider a vicinal surface
shown in \fig{fig1}, for regions with size
comparable to
or larger  than the step spacing, symmetry arguments can be
invoked to show that the average or homogenized displacements in such regions
vanish. Of course, on length scales much smaller than the step spacing there
can be local field fluctuations due to individual dipoles. However, at scales
larger than the step spacing, these fluctuations average out to zero and the
homogenized field is determined solely by  the spatial variations in the step
density.

\subsection{Local parts of the free energy}

 The local part of the free energy has contributions from two sources, one from
 the self-energy of  the force dipoles at the steps and the other due to the
 interaction of these dipoles with the epitaxial mismatch stress. In each of
 these cases, the interactions arise from the  local displacement fields near
 the steps as opposed to the homogenized displacement fields calculated in
 \eq{Discrdisp1}. Below, we consider each of these contributions separately.

 The self-energy of a force dipole distribution can be obtained from the interaction energy between force dipoles on
 a vicinal surface.
If we consider the vicinal surface in \fig{vicinalsurf}, the interaction energy of one of the dipoles with all the others can be expressed as
 \be
{E}_{dipole} = \frac{\pi^2}{6}\left[D_1^2\frac{\partial ^2}{\partial ^2 x}G_{11}(x\bE_1)\left.\right|_{x=1} + D_2^2\frac{\partial ^2}{\partial ^2 x}G_{22}(x\bE_1)\left.\right|_{x=1}\right] \frac{1}{d^2} \equiv \frac{G}{d^2},
\lab{Deself}
\ee
where $d= h_s/\tan \theta$ is the spacing between steps and the factor $\pi^2/6$
comes from evaluating the infinite sum $\sum_{i=1}^{\infty}{i^{-2}}$. The self-energy of a continuous distribution of dipoles can be calculated using \eq{Deself} as
\be
E_{self} = \int_{S_1} \frac{G}{d^2(x_1)} \frac{|h^{\prime}(x_1)|}{h_s} dx_1 = \int_{S_1} \frac{G}{h_s^3} |h^{\prime}(x_1)|^3 dx_1,
\lab{eself}
\ee
where the $local$ step spacing $d(x_1)$ is expressed in terms of the surface slope as $d(x_1) = h_s/|h^{\prime}(x_1)|$. The reason that the local approximation
works  for the self-energy is that the inverse square decay  of interaction between dipoles guarantees that
the contribution of steps that are far from a given step are much smaller in
magnitude than the contributions from those that are close by.  In distinct contract, the
logarithmic nature of the interaction between force monopoles
(given in \eq{Surfacetrac}) that arise from the spatial variation of the
dipoles, requires us to assign equal importance to steps that are both near
and far.  Hence, this contribution must be treated in the non-local approximation.

\begin{figure}[h]
\begin{center}
\includegraphics[width=5cm]{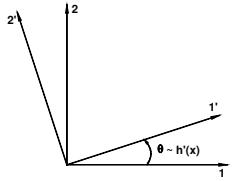}
\vspace{-3.5cm}
\parbox{1.0cm}{\ }\
\parbox{14cm}{
\caption{\lab{transformf}Transformation of coordinate axes to calculate
the interaction energy of the force dipoles and mismatch strain. The
coordinate axes in the transformed system are chosen so that the
$1^{\prime}$-axis lies along the vicinal surface and the $2^{\prime}$-axis
coincides with the normal to the vicinal surface.}}
\end{center}
\end{figure}

 The coupling of the strain field of a step with the mismatch strain gives rise
 to another local contribution to the free energy, which can also be evaluated
 by considering the vicinal surface in \fig{vicinalsurf}. The interaction energy between
 a positive step
and the mismatch strain is  $D_1 u_{1^{\prime},1^{\prime}}
+ |D_2| u_{2^{\prime},1^{\prime}}$,
where $u_{1^{\prime},1^{\prime}}$ and $u_{1^{\prime},2^{\prime}}$ are
displacement gradients in the transformed coordinate system shown in
\fig{transformf}. The coordinate axes in this transformed system are chosen
so that the $1^{\prime}$-axis lies along the vicinal surface and the
$2^{\prime}$-axis coincides with the normal to the vicinal surface. Using
the standard rules for transformation of strains, we have
\begin{eqnarray}
u_{1^{\prime},1^{\prime}} &=& \epsilon^{0}_{11} +  2\epsilon^{0}_{12}\theta + O(\theta^2)\nonumber \\
u_{2^{\prime},1^{\prime}} &=&  u^{0}_{2,1} + (\epsilon^{0}_{22}-\epsilon^{0}_{11})\theta + O(\theta^2).
\end{eqnarray}
For a continuous surface, the total contribution to the local part of the free
energy can be obtained by summing the interaction energies of each of the steps on
the surface with the mismatch strain, or
\begin{equation}
E_{dipole-strain} = \int_S \left[ D_1\frac{|h^{\prime}(x_1)|}{h_s} (\epsilon^{0}_{11} +  2\epsilon^{0}_{12} h^{\prime}(x_1))+  |D_2| \frac{h^{\prime}(x_1)}{h_s}(u^{0}_{2,1}+(\epsilon^{0}_{22}-\epsilon^{0}_{11})h^{\prime}(x_1))\right] dx_1.
\lab{elocal}
\end{equation}
Making use of the fact that $D_1 = \tilde{\beta_1}h_s$ and $|D_2| = \tau_0h_s$, it can be verified that the terms of $O(h^{\prime})$ and
$O(h^{\prime 2})$ in \eq{Continuume} add up to the local contribution evaluated in \eq{elocal} save the factor proportional to $u^{0}_{2,1}$. We do not have an explanation for this difference.

 Within the continuum description the parameter $\tilde{\beta}_1$ was introduced
 to include the effect of surface stretch  on the step formation energy.
 Comparing the continuum description with the discrete step models, we conclude
 that  this parameter is closely related to the intrinsic dipole at the
 step. If there are significant modifications in atomic bonding at the step
 edges, the intrinsic dipole will be large in magnitude. For such steps, strain
 can significantly reduce their formation energy. Atomistic calculations on
 reconstructed Si steps (Xie et al. 1994, Roland 1995) have shown large reduction in step
 formation energies due to strain. The results, however, were neither
 interpreted in terms of the intrinsic dipole nor used in a continuum framework
 to describing evolution of surfaces. Our continuum formalism accounts this
 crucial effect in the description of evolution of strained material surfaces.

\section{Energetics of strained islands}

\subsection{Continuum description}

We will now apply the continuum model to calculate the energy of strained islands. The island energies will
also be obtained directly by summing up the interactions of individual steps. It will be shown that
the continuum model not only provides an easy way of calculating the energetics but also
provides a direct of identifying the contributions to the total energy
arising from edges and corners. As we will see, extracting these quantities from the discrete step picture is
much more involved and tedious.

\begin{figure}[ht]
\begin{center}
\includegraphics[width=8cm]{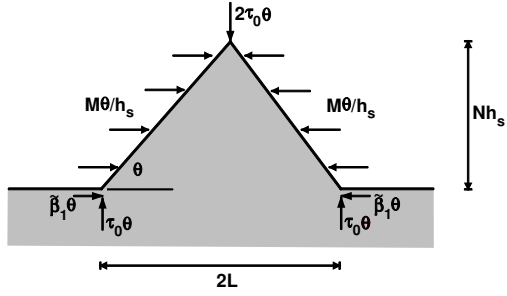}
\vspace{-6cm}
\parbox{1.0cm}{\ }\
\parbox{14cm}{
\caption{\lab{island}Two dimensional epitaxial island made up of $N$ positive
and negative steps and with side walls oriented angle $\theta$. The various
components of surface tractions that are needed to calculate the stress fields
in the island and substrate are shown in the figure.}}
\end{center}
\end{figure}

Let us consider the island made up of $N$ positive
and negative steps with base width $2L$ as shown in \fig{island};  the angle made by the
facet is given by $\theta \approx Nh_s/L$ and the two-dimensional volume
of the island is $A=LNh_s$. For the sake of simplicity, the surface Green's
function of an isotropic material will be employed in evaluating the
non-local contribution to the energy of the island.

Within the continuum description, the local contribution to energy of the island is
\be
E_L = 2A^{1/2}\theta^{1/2}\left[(\beta_1 + \tilde{\beta}_1\epsilon^0_{11}) + \tau_0(\epsilon^0_{22}-\epsilon^0_{11})\theta + \beta_3\theta^2\right] ,
\lab{islandl}
\ee
while the non-local contributions to the energy can be calculated from the surface tractions given in \eq{Surfacetrac}.
Since the slope on each of the side-walls of the island is a constant, the components of the surface tractions are given by
\begin{eqnarray}
f_1 &=& \frac{M}{h_s}\theta\left[\Theta(x-L)+\Theta(x+L)-2\Theta(x)\right] + \tilde{\beta}_1\theta\left[\delta(x-L)-\delta(x+L)\right], \nonumber \\
f_2 &=& \tau_0\theta[-2\delta(x)+\delta(x-L)+\delta(x+L)]
\label{islandtrac}
\end{eqnarray}
where $\Theta (\ ) $ is the unit step function and $\delta (\ ) $ is the Dirac delta function.
The first term in \eq{islandtrac}$_1$ comes from the force monopoles at the steps,
while the other terms that act at the apex and the edges of the island, as
shown in \fig{island}, come from the force dipoles at the step. The different
components of the non-local part of the island energy can be written as
\begin{eqnarray}
E_{MM} &=& \frac{\alpha M^2  \theta^2}{h_s^2} \left[ 2 \int_0^L (L-x) \log x dx - \int_{-L}^{L}(L-|x|)\log (L + x)dx \right] = -\alpha A \theta (C_{11ij}\epsilon_{ij}^0)^2 \log 4 , \nonumber \\
E_{MD} &=& -\frac{2 \alpha M \tilde{\beta}_1  \theta^2}{h_s} \int_0^L \log (2L/x-1) dx = - 2 \alpha \tilde{\beta}_1 A^{1/2} \theta^{3/2} (C_{11ij}\epsilon_{ij}^0) \log 4, \nonumber \\
E_{DD} &=& -\alpha {\tilde{\beta}_1}^2 \theta^2 \log(2L/h_s) -\alpha{\tau_0}^2 \theta^2 [4 \log (L/h_s) - \log(2L/h_s)]  \nonumber \\
&\approx& -\left[\alpha {\tilde{\beta}_1}^2 + 3\alpha {\tau_0}^2\right] \theta^2  \log(2L/h_s),
\lab{islandnl}
\end{eqnarray}
where the subscripts $M$ and $D$, respectively, indicate terms arising from
the monopoles and dipoles at the steps, and $\alpha = \frac{2(1-\nu^2)}{\pi E}$.

 The relative importance of the {\em edge} contribution to the island energy, given
 by \eq{islandnl}$_2$ plus \eq{islandnl}$_3$, can be determined by comparing its
 magnitude with the local contribution given by \eq{islandl}. It can easily
 be verified that the ratios
$E_{MD}/E_L$ and $E_{DD}/E_L$ are equal to $\theta$ and
$\log(\tilde{A})/\tilde{A}$ within factors of order unity, respectively,
where $\tilde{A} = A/h_s^2$. Since our focus is on shallow
islands ($\theta \ll 1$) whose dimensions are large compared to atomic dimensions,
it is clear that the edge contribution can be safely ignored and the energy of the island becomes
\be
E_{island} =  2A^{1/2}\theta^{-1/2}\left[(\beta_1 + \tilde{\beta}_1\epsilon^0_{11}) + \tau_0(\epsilon^0_{22}-\epsilon^0_{11})\theta + \beta_3\theta^2\right] -\alpha A \theta (C_{11ij}\epsilon_{ij}^0)^2\log 4.
\lab{eisland}
\ee
If the strain dependence of the first term is ignored, there is an energy
barrier to nucleation of epitaxial islands (Tersoff and Le Goues 1994). However
if the condition $\beta_1 + \tilde{\beta}_1\epsilon^0_{11} < 0$  is met, for
side-wall angles in the range $0 \le \theta \le \theta^*$ (see \fig{anisose} )
the nucleation barrier is absent. Indeed, experimental observations (Sutter and
Lagally 2000, Tromp et al. 2000) confirm that such shallow stepped mounds
islands form as a natural instability during strained heteroepitaxy.

\subsection{Discrete description}

The energy of the strained island in \fig{island} can be also obtained considering the interaction energy of discrete steps that make up the sidewalls of the islands. Using italicized symbols to denote energies, the local part of the energy that includes that step formation energy and the interaction of the force dipole with the mismatch strain can be expressed as
\be
{\cal{E}}_L = 2N\left[(\beta_1 + \tilde{\beta}_1\epsilon^0_{11}) + \tau_0(\epsilon^0_{22}-\epsilon^0_{11})\theta \right].
\lab{islandd}
\ee
If we compare this expression with the local contribution in the continuum description, we find that the self-energy of the
force dipoles does not appear in the discrete framework. The reason is that within the discrete model, both the self-energy
and the non-local part due the spatial variation of the step density have to be evaluated from the dipole-dipole interactions between the steps that constitute the island. The total interaction energy, which is a sum of interactions between pairs of steps can be written as
\be
{\cal{E}}_{int} = \frac{1}{2}\sum_{i,j}\left[\left(-M_iM_j-(M_i-M_j)D_1\frac{\partial}{\partial x_{ij}}+D_1^2\frac{\partial ^2}{\partial ^2 x_{ij}}\right)G_{11}(x_{ij}\bE_1) + D_2^2\frac{\partial ^2}{\partial ^2 x_{ij}}G_{22}(x_{ij}\bE_1) \right],
\label{Discretee}
\ee
where the steps are labeled with indices $i$ and $j$ and $x_{i,j} = x_i-x_j$. Below, we will evaluate the total energy for the island in \fig{island} by using the elastic Green's function of the isotropic solid.

 The double sum in \eq{Discretee} can be simplified by rearranging the terms in \eq{Discretee} (Shenoy et al. 1998, 2000), so that sum over one of the indices can be conveniently evaluated; the step interaction energies are then given by
\begin{eqnarray}
{\cal{E}}_{MM}&=& \alpha M^2  \left[ 2 \sum_{i=1}^{N-1} (N-i) \log i - \sum_{-N+1}^{N-1}(N-|i|)\log (N + i) \right], \nonumber \\
{\cal{E}}_{MD} &=& -2\alpha M \tilde{\beta}_1 \theta  \sum_{i = -N+1}^{N-1}\frac{N-|i|}{N + i}, \nonumber \\
{\cal{E}}_{DD} &=& \alpha {\tilde{\beta}_1}^2 \theta^2\sum_{i=1}^{2N-1}\frac{2N-i}{i^2}  + \alpha^2 \tau_0^2 \theta^2 \left[2\sum_{i=1}^{N-1}\frac{N-i}{i^2}+\sum_{i=-N+1}^{N-1}\frac{N-|i|}{(N + i)^2}\right].
\lab{Discrsum}
\end{eqnarray}
When $N$ becomes large, the discrete sums in \eq{Discrsum}$_1$ and \eq{Discrsum}$_2$
can be evaluated by converting to integrals.
The results so obtained are identical to the continuum results in \eq{islandnl}.
The dipole-dipole term  given by \eq{Discrsum}$_3$, in the limit of large $N$, yields
\be
{\cal{E}}_{DD} = \frac{\alpha \theta^2\pi^2}{3}[{\tilde{\beta}_1}^2 + \tau_0^2]   -\left[\alpha {\tilde{\beta}_1}^2 + 3\alpha {\tau_0}^2\right] \theta^2  \log(2L/h_s).
\ee
The first term is identical to the self-energy term of the dipoles in the continuum
description, while the next two terms agree with the
non-local contribution given in \eq{islandnl}$_3$. Thus, the expressions for energy of the island
computed using both the discrete and continuum
descriptions are in agreement.

The above exercise shows that the continuum model provides a direct route to computing the energy of
stepped surfaces without recourse to tedious evaluation of discrete sums. It also provides a direct way of looking at the
energies associated with corners and edges. It is difficult to obtain these contributions form the
discrete approach because the self-energy and the long-range part of the
dipole interactions are not clearly separated.

\section{Convergence of the variational formulation}

The variational formulation is based upon the expression of the surface shape in
terms of a Fourier series with a finite number of expansion coefficients; refer to
\eq{height}. In this appendix we investigate the convergence of
the expansion, focusing on the growth a single Fourier mode. The
goal of this exercise is to show that the amplitudes of the modes with small
wavelengths do not grow without bound.

In order to perform the stability analysis of short wavelength modes, the highly
nonlinear nature of the evolution equations behooves us to consider not just
the leading order contribution of the energy represented by the negative step
formation energy, but also the repulsive step
interaction energy.   It can be  shown that the nonlocal contribution,
given by the last term in \eq{Continuume}, becomes significant only when the
wavelengths are large. If only the second and third terms in \eq{Continuume} are
retained, the evolution of the surface shape consisting of a single mode,
\be
h(x) = a_n(t)\cos(nkx),
\ee
can be written as
\be
\dot{a}_n(t) = - \frac{4c  n^2 k^3}{\pi} \left[\beta_1 + \tilde{\beta}_1 \epsilon_{11}^0+2\beta_3a_n^2k^2n^2\right]\mathrm{Sgn}[a_n].
\label{modegrowth}
\ee
 If $\beta_1 + \tilde{\beta}_1\epsilon_{11}^0 < 0$, the growth of this mode is
 determined by a competition between the negative step formation energy, which
 promotes the growth of  small wavelength modes, and the repulsive step
 interactions. It is evident that $a_n$ achieves the stable amplitude
\be
|a_n| = \frac{1}{nk}\sqrt{\frac{-(\beta_1 + \tilde{\beta}_1 \epsilon_{11}^0)}{2 \beta_3}}.
\label{stableamp}
\ee
Because $|a_n|$ vanishes as $n \rightarrow \infty$, the use of a finite number
of terms in the expansion of surface shape in \eq{height} is reasonable.
This expectation is borne out by numerical experiments in the full problem.

In the above analysis, we did not include the effects of the substrate. Since
the mismatch strain is present only in the deposited film, the step formation
energy in the substrate is always positive. This implies that the amplitude
obtained in \eq{stableamp} represents an upper bound of sorts; if the effect of
the substrate is included, the stable amplitude would be smaller.

\section*{References}

\begin {list} {$-$}{\itemsep -0.1cm}{\partopsep -0.1cm}
\item R.~J.~Asaro and W.~A.~Tiller, Interface morphology development during stress corrosion cracking: part I. via diffusion, {\it Metall. Trans.} {\bf 3}, 1789-1796 (1972).
\item J.~A.~Floro, M.~B.~Sinclair, E.~Chason, L.~B.~Freund, R.~D.~Twesten, R.~Q.~Hwang and G.~A.~Lucadamo, Novel SiGe island coarsening kinetics: Ostwald ripening and elastic interactions, {\it Phys. Rev. Lett.} {\bf 84}, 701-704 (2000).
\item L.~B.~Freund, Evolution of waviness of the surface of a strained elastic solid due to stress-driven diffusion, {\it Int. J. Solids Structures} {\bf 32}, 911-923 (1995).
\item L.~B.~Freund, A surface chemical potential for elastic solids, {\it J.~Mech.~Phys.~Sols.} {\bf 46}, 1835-1844 (1998).
\item H.~Gao and W.~D.~Nix, Surface Roughening of heteroepitaxial thin films, {\it Annu. Rev. Mater. Sci.} {\bf 29}, 173-209 (1999).
\item M.~A.~Grinfeld, Instability of the separation boundary between non-hydrostatically stressed elastic solid and melt, {\it Sov. Phys. Dokl.} {\bf 31}, 831-834 (1996).
\item C.~Herring,  The use of classical macroscopic concepts in surface energy problems,  {\it Structure and Properties of Solid Surfaces}, edited by R. Gomer and C. S. Smith, University of Chicago Press, 5-72 (1953).
\item V.~I.~Marchenko and A.~Ya.~Parshin, Elastic properties of crystal surfaces, {\it Sov. Phys. JETP} {\bf 52}, 129-131 (1980).
\item G.~Medeiros-Ribeiro, A.~M.~Bratkovski, T.~I.~Kamins, D.~A.~A.~Ohlberg and R.~S.~Williams, Shape transition of Ge nanocrystals on
Si (001) surface from pyramids to domes, {\em Science} {\bf 279}, 353-355 (1998).
\item W.~H.~Press, S.~A.~Teukolsky and W.~T.~Vetterling and B.~P.~Flannery, "Numerical recipes: The art of scientific computing", (Cambridge University Press, Cambridge, 1992) pages 708-715.
\item A.~Rastelli, M.~Kummer and H.~von Kanel, Reversible shape evolution of Ge islands on Si(001), {\em Phys. Rev. Lett.} {\bf 87}, 6101-6104 (2001).
\item C.~Roland, Effect of stress on step energies and surface roughness, {\it MRS Bulletin} {\bf 21}, 27-30 (1996).
\item V.~B.~Shenoy, S.~Zhang and W.~F.~Saam, Bunching transitions on vicinal surfaces and quantum-$n$ mers, {\it Phys. Rev. Lett.} {\bf 81}, 3475-3478 (1998).
\item V.~B.~Shenoy, S.~Zhang and W.~F.~Saam, Step bunching transitions on vicinal surfaces with attractive step interactions, {\it Surf. Sci.} {\bf 467} 58-84 (2000).
\item D.~J.~Srolovitz, On the stability of surfaces of stressed solids, {\it Acta. Metall.} {\bf 37}, 621-625 (1989).
\item Z.~Suo, Motion of microscopic surfaces in materials, {\it Adv. Appl. Mech.} {\bf 33}, 193-294 (1997).
\item P.~Sutter and M.~G.~Lagally, Nucleationless three-dimensional island formation in low-misfit heteroepitaxy, {\it Phys. Rev. Lett.} {\bf 84}, 4637-4640 (2000).
\item J.~Tersoff and F.~K.~Le Goues, Competing relaxation mechanisms in strained layers, {\it Phys. Rev. Lett.} {\bf 72}, 3570-3573 (1994).
\item J.~Tersoff, Y.~H.~Phang, Z.~Y.~Zhang and M.~G.~Lagally, Step-bunching instability of vicinal surfaces under stress, {\it Phys. Rev. Lett.} {\bf 75}, 2730-2733 (1995).
\item R.~M.~Tromp, F.~M.~Ross and M.~C.~Reuter, Instability-driven SiGe island growth, {\it Phys. Rev. Lett.} {\bf 84}, 4641-4644 (2000).
\item A.~Vailionis, B.~Cho, G.~Glass, P.~Desjardins, D.~G.~Cahill and J.~E.~Greene, Pathway to strain-driven two-dimensional to three-dimensional transition during growth of Ge on Si(001), {\it Phys. Rev. Lett.} {\bf 85}, 3672-3675 (2000).
\item Y.~H.~Xie, G.~H.~Gilmer, C.~Roland, P.~J.~Silverman, S.~K.~Buratto, J.~Y.~Cheng, E.~A.~Fitzgerald, A.~R.~Kortan, S.~Schuppler, M.~A.~Marcus and P.~H.~Citrin, Semiconductor surface roughness - dependence on sign and magnitude of bulk strain, {\it Phys. Rev. Lett.} {\bf 73}, 3006-3009 (1994).
\item Y.~W.~Zhang and A.~F.~Bower, Numerical simulations of island formation in a coherent strained epitaxial thin film system, {\it J.~Mech.~Phys.~Sols.} {\bf 47}, 2273-2297 (1998).

\end {list}

\end{document}